\documentclass[aps, prd, twocolumn, print, nofootinbib]{revtex4}
\usepackage{amsmath}
\usepackage{txfonts}
\usepackage{amssymb}
\usepackage{mathrsfs}
\usepackage{graphicx}
\usepackage{epsfig}
\usepackage{dcolumn}
\usepackage{bm}

\begin{document}

\title{What can PSR J1640-4631 tell us about the internal physics of this neutron star?}
\author{Quan Cheng$^{1,2}$}\email{qcheng@ihep.ac.cn}
\author{Shuang-Nan Zhang$^{2,4}$}\email{zhangsn@ihep.ac.cn}
\author{Xiao-Ping Zheng$^{3,5}$}\email{zhxp@mail.ccnu.edu.cn}
\author{Xi-Long Fan$^{1,5}$}\email{fanxilong@outlook.com}

\affiliation{$^1$School of Physics and Technology, Wuhan University,
Wuhan 430072, China\\$^2$Key Laboratory of Particle Astrophysics,
Institute of High Energy Physics, Chinese Academy of Sciences,
Beijing 100049, China\\$^3$Institute of Astrophysics, Central China
Normal University, Wuhan 430079, China\\$^4$University of Chinese
Academy of Sciences, Beijing 100049, China\\$^5$School of Physics
and Mechanical \& Electrical Engineering, Hubei University of
Education, Wuhan 430205, China}

\date{Feb 2019}

\begin{abstract}
Gravitational wave emissions (GWEs) of pulsars could not only make
them promising targets for continuous gravitational wave searches
but also leave imprints in their timing data. We interpret the
measured braking index of PSR J1640-4631 with a model involving both
the GWE and dipole magnetic field decay. Combining the timing data
of PSR J1640-4631 and the theory of magnetic field decay, we propose
a new approach of constraining the number of precession cycles,
$\xi$, which is highly uncertain currently but can be tightly
related to the interior physics of a neutron star and its GWE. We
suggest that future observation of the tilt angle $\chi$ of PSR
J1640-4631 would not merely help to constrain $\xi$ but also
possibly provide information about the internal magnetic field
configuration of this pulsar. We find that $\xi$ would be larger
than previous estimates unless a tiny angle $\chi\lesssim5^\circ$ is
observed. Furthermore, a measured angle $\chi\gtrsim12^\circ$ would
indicate $\xi\gtrsim 10^6$, which is at least ten times larger than
that suggested previously.
\end{abstract}

\maketitle

\section{INTRODUCTION} \label{S:intro}
The braking indices of pulsars are indicative of the spin-down
mechanisms of neutron stars (NSs), which can be related to various
aspects of NS physics. Traditional scenarios of a rotating magnetic
dipole in vacuo show that pulsars should have braking indices $n=3$
(e.g., \cite{Ostriker:1969}). However, this simple model is
inconsistent with the observations of braking indices for all nine
young pulsars, of which eight pulsars have $n<3$ (see
\cite{Lyne:2015} and references therein) and only one has $n>3$
\cite{Archibald:2016}. To explain the $n<3$ braking indices, several
models have been invoked, including accretion of the fallback disc
around a NS \cite{Menou:2001}, braking torques due to relativistic
particle winds and magnetic dipole radiation (MDR) \cite{Xu:2001},
spin-down caused by quantum vacuum friction and MDR
\cite{Coelho:2016}, a decrease in the effective moment of inertia of
a NS as its interior normal matter becomes superfluid
\cite{Ho:2012a}, and an increase in the surface dipole magnetic
field due to either reemergence of the magnetic field buried after
birth \cite{Muslimov:1996} or evolution of the crustal magnetic
field \cite{Gourgouliatos:2015}.

The only young pulsar PSR J1640-4631 with $n>3$
\cite{Archibald:2016} observed hitherto\footnote{Recently, it has
been claimed that another young x-ray pulsar PSR J0537-6910 may have
$n=7$ as inferred from its complete timing data
\cite{Andersson:2017}. However, the result is inconclusive because
of frequent glitches of this pulsar.} has attracted great attention
and various models have been proposed to elucidate the large braking
index, for instance, magnetic dipole spin-down of a pulsar with a
plasma-filled magnetosphere \cite{Eksi:2016}, a combination of
dipole and wind braking \cite{Tong:2017}, spin-down of a
conventional NS (or even a exotic low-mass NS \cite{Chen:2016}) due
to MDR and gravitational wave emission (GWE) \cite{de
Araujo:2016a,de Araujo:2016c}, classical MDR braking but with dipole
field decay involved \cite{Gao:2017}. Theoretically, both GWE and
dipole field decay may be inevitable for a NS with a strong magnetic
field and an finitely conductive crust.

The strong magnetic fields of NSs could deform them into a quadruple
ellipsoid (see \cite{Glampedakis:2017} for a recent review), making
them promising sources for continuous GW searches using ground-based
GW detectors, such as advanced LIGO \cite{Abbott:2009}, Virgo
\cite{Acernese:2008}, and the planned Einstein Telescope
\cite{Punturo:2010}. Although no GW signals from known pulsars has
been detected during the first observing run of advanced LIGO
\cite{Abbott:2017}, the magnetically induced GWE could indeed affect
the spin evolution of NSs and leave some imprints in their timing
data. Moreover, for a deformed NS that is not in the minimum spin
energy state, to minimize its spin energy, free-body precession of
the star's magnetic axis around the spin axis will unavoidably
occur, which could lead to the change of the tilt angle between the
two axes.

Generally, the tilt angle evolution of a NS with a plasma-filled
magnetosphere \cite{Goldreich:1969} is determined by the MDR
\cite{Philippov:2014}, the GWE reaction \cite{Cutler:2000}, and
damping of the free-body procession due to internal dissipation
\cite{Alpar:1988,Cutler:2002}. Among them, the angle evolution
result from damping of the free-body precession can be related to a
critical parameter called the number of precession cycles, $\xi$
\cite{Jone:1976,Alpar:1988}. Since the damping mechanisms are not
clearly understood, only quite rough estimates for $\xi$ have been
proposed hitherto. For instance, as a possible damping mechanism,
Alpar \& Sauls \cite{Alpar:1988} studied the core-crust coupling due
to scattering of electrons off the neutron vortices and obtained
$\xi\approx 10^{2-4}$. On the other hand, damping of the stellar
free-body precession caused by elastic dissipation in the crust
gives a relatively large $\xi\lesssim 10^5$ \cite{Cutler:2002}.
However, this parameter is extremely important in discussing the GWE
of a NS (e.g., \cite{Stella:2005,Gualtieri:2011}), because $\xi$
could significantly affect the time scale over which the optimal
(unfavorable) configuration for GWE can be achieved, provided that
the star has a prolate (oblate) shape.

It has long been suggested that the dipole field that possibly
associated with the crustal field of a NS could decay due to Hall
drift and Ohmic dissipation (e.g.,
\cite{Jones:1988,Goldreich:1992}). The specific time scale for the
field decay is still uncertain, though typical time scales of
$\sim10^4$ yr (depending on the dipole field strength and the
density at the base of the crust) \cite{Cumming:2004,Dall'Osso:2012}
and $\sim10^6$ yr (depending on the electrical conductivity of the
crust) \cite{Goldreich:1992,JonesPB:2001,Ho:2011} were proposed for
Hall drift and Ohmic dissipation, respectively. Furthermore,
population synthesis studies of isolated radio pulsars suggested a
extremely long decay time scale of $\gtrsim 10^8$ yr if field decay
could indeed occur \cite{Mukherjee:1997}.

In this paper, we explain the braking index of PSR J1640-4631 based
on a model involving both GWE and dipole field decay, which are
natural consequences with the presence of strong magnetic fields of
a NS. We propose a new approach of estimating $\xi$ by using the
timing data of PSR J1640-4631 and the magnetic field decay theory.
We suggest that once the tilt angle of this pulsar is measured, we
could not only put constraints on the highly uncertain parameter
$\xi$ but also possibly know about its internal magnetic field
configuration. Interestingly, the value of $\xi$ would be larger
than previous results unless a tiny tilt angle ($\lesssim 5^\circ$)
is observed. The paper is organized as follows. The evolutionary
model for PSR J1640-4631 is presented in Sec. \ref{Sec II}. We
introduce the theory of magnetic field decay in Sec. \ref{Sec III}.
Our results are given in Sec. \ref{Sec IV}. Finally, a conclusion
and some brief discussions about possible physical explanations of a
large $\xi$ and its influence on the GWEs from newborn magnetars are
provided in Sec. \ref{Sec V}.

\section{EVOLUTION OF PSR J1640-4631}\label{Sec II}

Using the \textit{NuSTAR} X-ray observatory, Gotthelf \textit{et
al}. \cite{Gotthelf:2014} discovered the pulsar PSR J1640-4631,
whose period and first period derivative are $P=206$ ms and
$\dot{P}=9.758\times10^{-13}$ s/s, respectively. Recently, by
performing a phase-coherent timing analysis of the x-ray timing data
of PSR J1640-4631 observed with \textit{NuSTAR}, Archibald
\textit{et al}. \cite{Archibald:2016} obtained its second period
derivative and braking index $n=3.15(3)$.

For a pulsar with a corotating plasma magnetosphere
\cite{Goldreich:1969} that spins down mainly due to MDR and magnetic
deformation-induced GWE, its angular frequency evolution has the
following form \cite{Cutler:2000,Spitkovsky:2006}:
\begin{eqnarray}
\dot{\omega}=-\frac{2G\epsilon_{\rm B}^2I\omega^5}{5c^5}
\sin^2\chi(1+15\sin^2\chi)-\frac{kB_{\rm
d}^2R^6\omega^3}{Ic^3}(1+{\rm sin}^2\chi) \label{dwdt},
\end{eqnarray}
where $\epsilon_{\rm B}$ is the ellipticity of magnetic deformation,
$I$ the moment of inertia, $\chi$ the tilt angle, $k$ the
coefficient related to MDR, $B_{\rm d}$ the surface dipole magnetic
field at the magnetic pole, and $R$ the stellar radius. Hereinafter,
we adopt $k=1/6$, and take canonical values for the parameters of
the presumed $1.4M_\odot$ NS as $I=10^{45}$ g ${\rm cm}^2$ and
$R=10$ km.\footnote{We note that the value of $k$ is still in debate
(see Refs.
\cite{Spitkovsky:2006,Philippov:2014,Contopoulos:2014,Philippov:2015}).
However, adopting different values for $k$(=1/4) and $R$(=12 km)
could affect the value of $\xi$ by at most a factor of two.} We
define a ratio $\eta=\dot{\omega}_{\rm MDR}/\dot{\omega}_{\rm
GWE}=5kc^2B_{\rm d}^2R^6(1+\sin^2\chi)/[2G\epsilon_{\rm
B}^2I^2\omega^2(1+15\sin^2\chi)\sin^2\chi]$, where
$\dot{\omega}_{\rm MDR}$ and $\dot{\omega}_{\rm GWE}$ are the
MDR-induced and GWE-induced spin-down rate, respectively. Though the
GWE braking becomes maximal when $\chi=\pi/2$ is taken, one still
has $\eta\gg1$ for $\left|\epsilon_{\rm B}\right|\ll
8.69\times10^{-3}(B_{\rm d}/10^{13}~{\rm G})$, as $\omega$ is known
for PSR J1640-4631. We will show that no matter whether the internal
fields of this NS are poloidal-dominated (PD) or toroidal-dominated
(TD), the theoretically estimated $\epsilon_{\rm B}$ is far beneath
this limit.

Previous studies have shown that the NS equation of state, the
magnetic energy, the internal magnetic configuration, and the
presence of proton superconductivity in the core (which may change
the interior magnetic field distribution) could all affect the
magnetic deformation of a NS (e.g., Refs.
\cite{Haskell:2008,Dall'Osso:2009}). Lots of theoretical
calculations have been made to obtain the ellipticity (see, e.g.,
Refs.
\cite{Bonazzola:1996,Cutler:2002,Haskell:2008,Dall'Osso:2009,Ciolfi:2009,Ciolfi:2010,Gualtieri:2011,Mastrano:2011,Mastrano:2012,Lander:2013,Lasky:2013,Mastrano:2015,Dall'Osso:2015}).
For a young NS like PSR J1640-4631, its interior temperature is
probably lower than the critical temperature for proton
superconductivity \cite{Page:2014}, even if only modified Urca
cooling occurs \cite{Page:2006}. Hence, to estimate $\epsilon_{\rm
B}$ of PSR J1640-4631, the effect of proton superconductivity should
be involved, as that done in Ref. \cite{Lander:2013}.

After considering type-II proton superconductivity in the interior
of a NS, Lander \cite{Lander:2013} self-consistently obtained an
equilibrium configuration that consists of a mixed poloidal-toroidal
field and derived the corresponding magnetic ellipticity
\begin{eqnarray}
\epsilon_{\rm B}=3.4\times10^{-7}\left({B_{\rm d}\over10^{13}~{\rm
G}}\right)\left({H_{\rm c1}(0)\over10^{16}~{\rm G}}\right)
\label{epsil1},
\end{eqnarray}
where the central critical field strength is taken to be $H_{\rm
c1}(0)=10^{16}$ G \cite{Lander:2013}. In this field configuration,
since the dominant part is the poloidal component, the NS has a
oblate shape ($\epsilon_{\rm B}>0$). This configuration is partially
akin to the twisted-torus configuration found in numerical
simulations \cite{Braithwaite:2006}. The main difference is that in
the latter configuration, the toroidal field may be dominant
\cite{Braithwaite:2009}, the NS possibly has a prolate shape
($\epsilon_{\rm B}<0$). With type-II proton superconductivity
involved, and based on the twisted-torus configuration, a
calculation of $\epsilon_{\rm B}$ is presented in Ref.
\cite{Mastrano:2012}. However, the results are very rough and only
upper limits are given for $\epsilon_{\rm B}$ because the
superconducting stellar interior is assumed to have a homogeneous
magnetic permeability, which is in fact physically implausible.
Since there is no self-consistent calculations for the ellipticity
of a superconducting NS that has a TD twisted-torus field
configuration inside currently, we simply adopt $\epsilon_{\rm B}$
derived for the pure toroidal configuration as a substitution, which
takes the form \cite{Akgun:2008}
\begin{eqnarray}
\epsilon_{\rm B}\approx-10^{-8}\left({H\over10^{15}~{\rm
G}}\right)\left({\bar{B}_{\rm in}\over10^{13}~{\rm G}}\right)
\label{epsil2},
\end{eqnarray}
where $H\approx10^{15}$ G is the critical field strength and
$\bar{B}_{\rm in}$ the volume-averaged strength of the internal
toroidal field. It is generally hard to determine $\bar{B}_{\rm in}$
of a NS. Fortunately, the observed positive correlation between the
surface temperatures and dipole magnetic fields of isolated NSs
(with $B_{\rm d}\gtrsim 10^{13}$ G) indicates that strong toroidal
fields with volume-averaged strengths of $\sim10B_{\rm d}$ possibly
exist in NS crusts \cite{Pons:2007}. We thus assume that the
strengths of the crustal toroidal fields are representative of
$\bar{B}_{\rm in}$ of the whole stars, that is, $\bar{B}_{\rm
in}\simeq10B_{\rm d}$. Internal fields that are one order of
magnitude (or more) higher than dipole fields may indeed be present
in young pulsars (see Ref. \cite{Glampedakis:2012}).

It should be noted that the internal fields which determine the
ellipticity may also decrease as the star evolves. Here we assume
that the relation between the internal fields and $B_{\rm d}$
remains unchanged and the expression for $\epsilon_{\rm B}$ given by
Eq. (\ref{epsil1}) or (\ref{epsil2}) still holds with the decay of
$B_{\rm d}$, though a global long-term numerical simulation is
needed to reveal how internal fields and $\epsilon_{\rm B}$ vary
with time. Interestingly, a time-dependent $\epsilon_{\rm B}$, as
also considered in Ref. \cite{de Araujo:2016c}, can hardly change
our results in comparison with the case of a time-independent
$\epsilon_{\rm B}$. The reason is that adopting a time-dependent
$\epsilon_{\rm B}$ results in a factor $(1+1/\eta)\simeq 1$ just
before the term $\dot{B}_{\rm d}/B_{\rm d}$ in Eq. (\ref{bi2}),
which is 1 for the case of a time-independent $\epsilon_{\rm B}$.
From Eqs. (\ref{epsil1}) and (\ref{epsil2}), we can see that these
estimated $\epsilon_{\rm B}$ are consistent with the requirement of
$\eta\gg1$. The GWE braking can therefore be neglected due to its
little effect on the spin-down of PSR J1640-4631. However, the GWE
could still affect the pulsar's tilt angle evolution.

The tilt angle evolution of a magnetically deformed NS with a plasma
magnetosphere is given by
\cite{Cutler:2000,Jones:2001,Dall'Osso:2009,Philippov:2014}:
\begin{eqnarray}
\dot{\chi}=\left\{ \begin{aligned}
         -\frac{2G}{5c^5}I\epsilon_{\rm
B}^2\omega^4\sin\chi&\cos\chi(15\sin^2\chi+1)-{\epsilon_{\rm
B}\over\xi P}{\rm tan}\chi\\&-\frac{kB_{\rm d}^2R^6\omega^2}{Ic^3}\sin\chi\cos\chi,~{\rm for}~\epsilon_{\rm B}>0 \\
                  -\frac{2G}{5c^5}I\epsilon_{\rm
B}^2\omega^4\sin\chi&\cos\chi(15\sin^2\chi+1)-{\epsilon_{\rm
B}\over\xi P}{\rm cot}\chi\\&-\frac{kB_{\rm
d}^2R^6\omega^2}{Ic^3}\sin\chi\cos\chi,~{\rm for}~\epsilon_{\rm
B}<0.
                          \end{aligned} \right.
\label{dchi}
\end{eqnarray}
The first and third terms of the above formula represent the
alignment effects caused by the GWE and MDR, respectively. The
second term represents the angular evolution from damping of the
stellar free-body procession due to internal dissipation. Depending
on the shape of a NS (or the sign of $\epsilon_{\rm B}$), this
effect could either decrease or increase $\chi$. Actually, Eq.
(\ref{dchi}) stands for the main difference as compared to previous
models \cite{Chen:2016,de Araujo:2016a,de Araujo:2016c}, in which
these mechanisms for tilt angle evolution were not considered.

By taking both the field decay and tilt angle evolution into
account, the braking index reads
\begin{eqnarray}
n = &3-{2P\over\dot{P}}\left\{{\dot{B}_{\rm
d}\over B_{\rm d}}+\dot{\chi}\sin\chi\cos\chi\left[\frac{1}{1+\sin^2\chi} +\right.\right.\nonumber\\
&\phantom{=\;\;}\left.\left.\frac{1+30\sin^2\chi}{\eta\sin^2\chi\left(1+15\sin^2\chi\right)}
\right]\right\} \label{bi2},
\end{eqnarray}
where $\dot{B}_{\rm d}$ is the decay rate of $B_{\rm d}$. We will
see below Eq. (\ref{bi2}) is a critically link that relates $\xi$ in
Eq. (\ref{dchi}) to the timing data of PSR J1640-4631 and the field
decay time scale $\tau_{\rm D}=-B_{\rm d}/\dot{B}_{\rm d}$
determined by the field decay theory.

\section{THE THEORY OF MAGNETIC FIELD DECAY}\label{Sec III}

The decay rate of $B_{\rm d}$ is determined by the specific field
decay mechanisms, which are generally considered to be Hall drift
and Ohmic dissipation if the dipole field has a crustal origin.
However, the mathematical form of field decay is still not clearly
known. For simplicity, we consider two typical decay forms that
introduce the least parameters. The first one is the exponential
form \cite{Pons:2007,Dall'Osso:2012}
\begin{eqnarray}
{dB_{\rm d}\over dt}=-{B_{\rm d}\over \tau_{\rm D}} \label{dBdt1},
\end{eqnarray}
where $\tau_{\rm D}$ is the dipole field decay time scale. The
second one is the nonlinear form
\cite{Dall'Osso:2012,Ho:2012b,Gao:2017}
\begin{eqnarray}
{dB_{\rm d}\over dt}=-{B_{\rm d}\over {\tau_{\rm D}+t}}
\label{dBdt2},
\end{eqnarray}
where $t$ is the actual age of the pulsar. Generally, $\tau_{\rm D}$
may be determined by both Hall drift and Ohmic dissipation in the
crust as $1/\tau_{\rm D}=1/\tau_{\rm Hall}+1/\tau_{\rm Ohmic}$ (see,
e.g., \cite{Gao:2017}), where $\tau_{\rm Hall}$ and $\tau_{\rm
Ohmic}$ are Hall drift and Ohmic dissipation time scales,
respectively. It should also be noted that Hall drift itself is a
non-dissipative process, however, could substantially accelerate the
field decay by changing the large scale magnetic field into small
scale components, which would decay rapidly due to Ohmic dissipation
\cite{Goldreich:1992,Muslimov:1994}. In this case, the field decay
time scale may be set by the Hall time scale in the crust as
$\tau_{\rm D}=\tau_{\rm Hall}\simeq1.2\times10^4(10^{15}~{\rm
G}/B_{\rm d})~{\rm yr}$ \cite{Cumming:2004,Dall'Osso:2012}.

Furthermore, if Ohmic dissipation dominates the crustal field decay
process, as indicated by the positive correlation between the
surface temperatures and dipole fields of isolated NSs
\cite{Pons:2007}, the dipole fields which are assumed to be
proportional to the crustal fields may decay on the same time scale
$\tau_{\rm D}=\tau_{\rm Ohmic}\simeq5\times10^5$ or $10^6$ yr as the
latter \cite{Pons:2007}. Lastly, numerical modeling of the coupled
magnetic field evolution in the crust and the core of a NS shows
that $B_{\rm d}$ could decay over a time scale $\tau_{\rm D}\simeq
150$ Myr due to the combined effects of flux tube drift in the core
and Ohmic dissipation in the crust \cite{Bransgrove:2018,Zhu:2018}.
This may represent the longest field decay time scale predicted
theoretically, and it is also consistent with the results of pulsar
population synthesis \cite{Mukherjee:1997}.

In Fig. \ref{Fig1} we show $\tau_{\rm D}$ as a function of $\chi$.
The latter is related to $B_{\rm d}$ via Eq. (\ref{dwdt}) by
neglecting the term of GWE. From the timing data of PSR J1640-4631,
we obtain $B_{\rm d}\sim2\times10^{13}$ G. Thus $\tau_{\rm
Hall}(\chi)$ (black solid line) is approximately equal to $\tau_{\rm
Ohmic}\simeq5\times10^5$ yr (black dashed line). If $\tau_{\rm
D}(\chi)$ follows the form $\tau_{\rm D}(\chi)=1/[1/\tau_{\rm
Hall}(\chi)+1/\tau_{\rm Ohmic}]$, its minimum value at $\chi$ can be
obtained by taking $\tau_{\rm Ohmic}=5\times10^5$ yr, as shown by
the black dash-dot-dotted line (also the lower boundary of the blank
region) in Fig. \ref{Fig1}. A larger $\tau_{\rm Ohmic}$ can shift
this boundary upwards, but should not surpass $\tau_{\rm
Hall}(\chi)$. The maximum value of $\tau_{\rm D}(\chi)$ at $\chi$
could be determined by $\tau_{\rm Ohmic}$, which may be
$5\times10^5$, $10^6$ (black dotted line), or $1.5\times10^8$ yr
(black dash-dotted line) if Ohmic dissipation dominates the field
decay.\footnote{Here we attribute $\tau_{\rm D}(\chi)\simeq150$ Myr
to the effect of crustal Ohmic dissipation but keep in mind that
flux tube drift in the core region also plays an important role.}
The upper boundary of the blank region in Fig. \ref{Fig1}
corresponds to $\tau_{\rm D}(\chi)=1.5\times10^8$ yr, above which
should be excluded following the field decay theory.

From Eqs. (\ref{dBdt1}) and (\ref{dBdt2}), we have $\tau_{\rm
D}=-B_{\rm d}/\dot{B}_{\rm d}$ and $\tau_{\rm D}=-B_{\rm
d}/\dot{B}_{\rm d}-t$, respectively. The actual age $t$ of PSR
J1640-4631 remains unconstrained from observations currently, though
an estimate of $t\sim3000$ yr (close to its characteristic age
$\tau_{\rm c}=3350$ yr \cite{Gotthelf:2014}) was proposed on basis
of the dipole field decay \cite{Gao:2017}. Assuming
$t\simeq\tau_{\rm c}$, from Fig. {\ref{Fig1}} we can see that $t$ is
far below the lower boundary of $\tau_{\rm D}(\chi)$. Therefore,
hereinafter we can safely neglect the term $t$ and determine the
decay time scale via $\tau_{\rm D}=-B_{\rm d}/\dot{B}_{\rm d}$.

\section{RESULTS}\label{Sec IV}

\begin{figure}
\resizebox{\hsize}{!}{\includegraphics{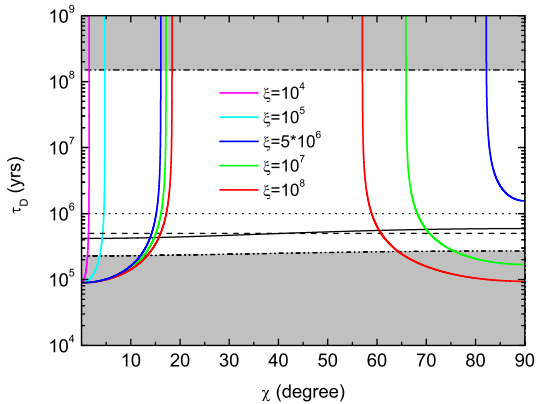}} \caption{Dipole
field decay time scale $\tau_{\rm D}$ versus tilt angle $\chi$. This
figure shows a comparison between $\tau_{\rm D}(\chi)$ derived using
timing data (colored lines) and that obtained based on the magnetic
field decay theory (black lines). The NS is assumed to have PD
internal fields. See the text for details.} \label{Fig1}
\end{figure}

By substituting the observed $P$, $\dot{P}$, $n=3.15$, and Eq.
(\ref{dchi}) into Eq. (\ref{bi2}), and taking $\xi$ as a free
parameter, one can solve for $\tau_{\rm D}=-B_{\rm d}/\dot{B}_{\rm
d}$ versus $\chi$. The evolution curves $\tau_{\rm D}(\chi)$ for
different $\xi$ are shown by the colored curves in Fig. \ref{Fig1}.
Since the evolution of $\chi$ depends on the shape of the NS, in
Fig. \ref{Fig1}, we first show the results for the PD case with
$\epsilon_{\rm B}$ given by Eq. (\ref{epsil1}).
\begin{figure}
\resizebox{\hsize}{!}{\includegraphics{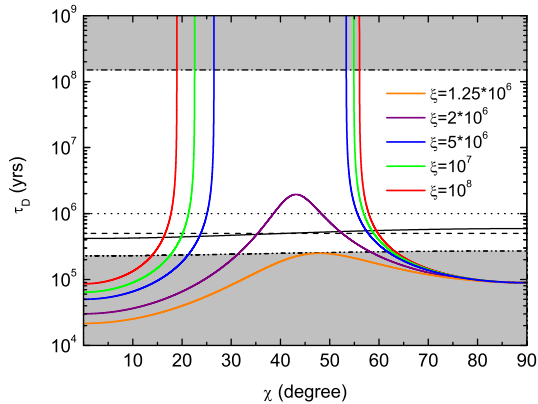}} \caption{The same
as Fig. \ref{Fig1}, however, the NS is assumed to have TD internal
fields. See the text for details.} \label{Fig2}
\end{figure}

The constraint on $\xi$ is set by the fact that at a certain $\chi$,
$\tau_{\rm D}(\chi)$ derived from timing data of PSR J1640-4631
should be equal to $\tau_{\rm D}(\chi)$ obtained based on the field
decay theory. That is, it requires that the colored curve should at
least intersect with one of the black curves, as presented in Fig.
\ref{Fig1}. If the internal fields of this pulsar are PD, for the
number of precession cycles in a wide range of $10^4\lesssim
\xi\lesssim 10^8$, each of the colored curves has at least one
intersection with the black lines. The interactions are distributed
within $2^\circ\lesssim\chi\lesssim18^\circ$ and
$57^\circ\lesssim\chi\lesssim90^\circ$. Specifically, for
$\xi\lesssim 10^5$, all the intersections are within $\chi\lesssim
5^\circ$. For $5\times10^6\lesssim \xi\lesssim 10^8$, $\tau_{\rm
D}(\chi)$ derived via Eq. (\ref{bi2}) splits into two branches, of
which the left one has interactions at
$12^\circ\lesssim\chi\lesssim18^\circ$, and the right one has
interaction(s) at $57^\circ\lesssim\chi\lesssim82^\circ$. Even if
$\xi\gtrsim10^9$ (which might be unphysical) is taken, no
interactions could be found for intermediate angles
$18^\circ\lesssim\chi\lesssim57^\circ$.

We also investigate another possibility that this NS has TD internal
fields with $\epsilon_{\rm B}$ given by Eq. (\ref{epsil2}). The
results are presented in Fig. \ref{Fig2}, which shows that in order
to have at least one intersection between the curve $\tau_{\rm
D}(\chi)$ obtained based on the timing data and the black
dash-dot-dotted line, the lower limit for the number of precession
cycles can be set as $\xi\gtrsim 1.25\times10^6$ (the orange curve).
All the intersections are distributed within $14^\circ\lesssim
\chi\lesssim 63^\circ$ for $1.25\times10^6\lesssim\xi\lesssim10^8$.
For the tilt angle in the ranges $\chi\lesssim 14^\circ$ and
$\chi\gtrsim 63^\circ$, there is no intersections even though an
(unphysically) large $\xi\gtrsim10^9$ is adopted. The same as in the
PD case, $\tau_{\rm D}(\chi)$ derived from the timing data also
shows a bifurcation for $5\times10^6\lesssim \xi\lesssim 10^8$.

Therefore, we suggest that future observations of the tilt angle of
PSR J1640-4631 would probably help to probe its internal magnetic
field configuration and put constraints on the number of precession
cycles. For instance, a small measured angle $\chi\lesssim 14^\circ$
possibly supports a PD internal field configuration because no
intersections are found for $\chi$ in this range in the TD case.
Moreover, a small value for the number of precession cycles
$\xi\lesssim 10^5$ as suggested in previous work
\cite{Alpar:1988,Cutler:2002,Jones:2001,Gualtieri:2011} could be
confirmed only if a tiny angle $\chi\lesssim 5^\circ$ is observed.
Beyond this angle, $\xi$ would be larger than previous estimates no
matter whether the internal fields are PD or TD. With some more
calculations we find that as long as an angle $\chi\gtrsim12^\circ$
is observed,\footnote{This is the largest lower limit required to
satisfy $\xi\gtrsim 10^6$, which is derived for the PD case and by
taking $\tau_{\rm D}(\chi)=150$ Myr.} one would have $\xi\gtrsim
10^6$, irrespective of the internal field configuration. A large
angle $\chi\gtrsim 63^\circ$ may also indicates the PD scenario,
however, the required $\xi$ is in the range
$10^6\lesssim\xi\lesssim10^8$, at least $\sim10-\-10^3$ larger than
previous results. In contrast, an intermediate angle
$18^\circ\lesssim\chi\lesssim57^\circ$ seems to favor a TD internal
field configuration, and a large $\xi$ whose lower limit is
$1.25\times10^6$. Only for the measured angle in two small ranges
$14^\circ\lesssim\chi\lesssim18^\circ$ and
$57^\circ\lesssim\chi\lesssim63^\circ$, we could not deduce whether
the poloidal or the toroidal field is dominant in the NS interior.

\section{CONCLUSION AND DISCUSSIONS}\label{Sec V}

Based on the timing data of PSR J1640-4631 and the magnetic field
decay theory, we propose a new method of estimating a vital but
presently highly unknown parameter called the number of precession
cycles, $\xi$. In the modeling, we considered different internal
magnetic field configurations, field decay formulas, and field decay
time scales. We conclude that if the tilt angle $\chi$ of PSR
J1640-4631 could be measured through polarization observation using
future x-ray telescopes (e.g., eXTP \cite{Zhang:2016}), we may get
quite valuable information about $\xi$ and the internal magnetic
fields of this pulsar. Most importantly, irrespective of the
internal field configuration, as long as the angle is observed to be
$\chi\gtrsim5^\circ$, $\xi$ should be constrained to be larger than
previous results
\cite{Alpar:1988,Cutler:2002,Jones:2001,Gualtieri:2011}. As a
conservative estimate, a measured angle $\chi\gtrsim12^\circ$ would
indicate $\xi\gtrsim 10^6$, which is at least ten times larger than
that suggested previously.

Physically, a large $\xi$ indicates that some rather weak damping
mechanisms are responsible for the dissipation of the precessional
energy. In the crust, if phonon excitations govern the interactions
between vortices and lattices, the mutual friction parameter, whose
reciprocal is approximately equal to $\xi$, could be as large as
${\mathcal{B}}\approx10^{-8}$ (e.g.,
\cite{Haskell:2017,Haskell:2018}). Therefore, an inferred large
$\xi\approx10^8$ may suggest that most of the precessional energy is
dissipated in the crust due to vortex-lattice interaction controlled
by phonon excitations. On the other hand, in the core some
(\textit{unknown}) weak damping mechanisms rather than
electron-vortex interaction may be dominant, as recently found in
\cite{Haskell:2018} that in the core
${\mathcal{B}}\sim10^{-7}-10^{-6}$ is required to interpret the
rising processes of three large Crab glitches. If $\xi$ is
constrained to be large in the future, it would greatly expedite our
understanding of complex interactions in NSs.

Furthermore, a large $\xi$ means a long time scale for a prolate NS
(e.g., newborn magnetars) to achieve the orthogonal configuration
\cite{Stella:2005} provided that $\chi$ could not rapidly increase
during very early period \cite{Dall'Osso:2009}. Thus, if newborn
magnetars have a large $\xi$, their GWEs may be weak and not easy to
be detected.

Finally, though we only performed a case study for PSR J1640-4631,
we should stress that our new method of estimating $\xi$ also
applies to other eight pulsars with a measured braking index. The
derived constraints on $\xi$ for these pulsars may be different from
that for PSR J1640-4631. This is reasonable because for different
pulsars the dominant interior interactions and the internal magnetic
field configurations are possibly various. A detailed analysis for
other pulsars will be presented in a subsequent paper.

\acknowledgements We thank the anonymous referees, W. C. G. Ho, and
D. I. Jones for helpful comments and suggestions. Quan Cheng
acknowledges funding support by China Postdoctoral Science
Foundation under grant No. 2018M632907. This work is also supported
by the National Natural Science Foundation of China (Grants No.
11773011, No. 11373036, No. 11133002, No. 11673008, and No.
11622326), the National Program on Key Research and Development
Project (Grants No. 2016YFA0400802, and No. 2016YFA0400803), and the
Key Research Program of Frontier Sciences, CAS (Grant No.
QYZDY-SSW-SLH008).

\end{document}